\documentclass[twocolumn,           
               showpacs,            
               preprintnumbers,     
               aps,                 
               prl,          	    
               letterpaper,             
               superscriptaddress,      
               nofootinbib,         
               tightenlines,        
               floats,floatfix      
               ]{revtex4-1}
\usepackage{graphicx}  
\usepackage{dcolumn}   
\usepackage{bm}        
\usepackage{amsmath,amssymb}
\usepackage{hyperref}

\begin{document}

\title{The mass discrepancy-acceleration relation: a universal
  maximum dark matter acceleration and implications for the
  ultra-light scalar field dark matter model}
\author{L. Arturo Ure\~{n}a-L\'{o}pez} 
 \email{lurena@ugto.mx}
\affiliation{%
Departamento de F\'isica, DCI, Campus Le\'on, Universidad de
Guanajuato, 37150, Le\'on, Guanajuato, M\'exico.}
\author{Victor H. Robles}
\email{vrobles@fis.cinvestav.mx}
\affiliation{Department of Physics and Astronomy, University of
  California, Irvine, 4129 Frederick Reines Hall, Irvine, CA 92697,
  USA}
\author{T. Matos}  
\email{tmatos@fis.cinvestav.mx}
\affiliation{Departamento de F\'isica, Centro de Investigaci\'on y de Estudios Avanzados del IPN, A.P. 14-740, 07000 M\'exico D.F.,
  M\'exico.}

\date{\today}

\begin{abstract}
Recent analysis of the rotation curves of a large sample of galaxies
with very diverse stellar properties reveal a relation between the
radial acceleration purely due to the baryonic matter and the one
inferred directly from the observed rotation curves. Assuming the
dark matter (DM) exists, this acceleration relation is tantamount to
an acceleration relation between DM and baryons. This leads us to a
universal maximum acceleration for all halos. Using the latter in DM
profiles that predict inner cores implies that the central surface
density $\mu_{DM} = \rho_s r_s$ must be a universal constant, as
suggested by previous studies in selected galaxies, revealing a strong
correlation between the density $\rho_s$ and scale $r_s$ parameters in
each profile. We then explore the consequences of the constancy of
$\mu_{DM}$ in the context of the ultra-light scalar field dark matter
model (SFDM). We find that for this model $\mu_{DM} = 648 \, M_\odot {\rm
  pc}^{-2}$, and that the so-called WaveDM soliton profile should be
an universal feature of the DM halos. Comparing with data from the
Milky Way and Andromeda satellites, we find that they are 
consistent with a boson mass of the scalar field particle of the order
of $10^{-21} \, {\rm eV}/c^2$, which puts the SFDM model in agreement
with recent cosmological constraints.
\end{abstract}

\maketitle


In the standard cold dark matter (CDM) paradigm, the dark matter (DM)
is $\sim$ 22\% of the total matter budget in the
Universe\cite{Aghanim:2016sns}, it is assumed to be collisionless and
non-relativistic after decoupling, forming structure hierarchically,
i.e. small halos merge to form more massive systems. Several
cosmological CDM simulations that exclude the luminous matter have
confirmed this formation scenario showing that in all scales halos
share a common density profile with a characteristic cusp (divergent
density) near their
centers\cite{Navarro:1996gj,Springel:2005nw,Dutton:2014xda,Onorbe:2015ija}. Assuming
galaxies are formed in these halos allows the comparison with
observations. However, detailed comparison with the dynamics of
low-mass galaxies have led to some longstanding discrepancies,
e.g. cusp-core problem, the satellite abundance,
too-big-to-fail\cite{deBlok:2009sp,Klypin:1999uc,BoylanKolchin:2011dk,Rodrigues:2017vto}. Galaxies
are then one of the greatest challenges for the CDM paradigm.

One quantity that summarizes the properties of the rotation curves in
galaxies is the so-called mass-discrepancy-acceleration relation
(MDAR)\cite{McGaugh:2016leg,McGaugh:2016oct}. The MDAR is observed for
a diverse sample of galaxies, from high to low surface brightness
galaxies and of different sizes and morphologies. This seemingly
independence of the relation on the luminous matter
suggests that if there is a common origin to the MDAR, it is probably
not strongly tied to the baryonic matter. There are two
straightforward approaches to explain the origin of the MDAR, one is
to assume that the acceleration relation results from modifying the
gravitational force as suggested by the MOND
hypothesis\cite{Milgrom:2016uye} (but see
also\cite{Diez-Tejedor:2016fdn}), and the second is that the MDAR is a
direct consequence of the intrinsic properties of the
DM\cite{Salucci:2016vxb}. The latter approach will be assumed for the
purposes of this Letter.

Considering the equivalent DM halo to explain the MDAR, it can be
shown that the gravitational acceleration of the DM can be found
from\cite{McGaugh:2016leg}
\begin{equation}
  \label{eq:11}
  g_h = \frac{g_{\rm bar}}{e^{\sqrt{g_{\rm bar}/g^\dagger}} -1} \, , 
\end{equation}
where $g_{\rm bar}$ is the acceleration produced by the baryons in the
galaxy and $g^\dagger = 1.2 \times 10^{-10} {\rm m} \, {\rm s}^{-2}$
is a characteristic acceleration obtained from fitting the
data. Apart from the characteristic acceleration $g^\dagger$, there exists 
a maximal acceleration $g_{h,{\rm max}}$ that can be obtained
from Eq.~\eqref{eq:11}, or from any other MOND
function\cite{Brada:1998mi}. Given that Eq.~\eqref{eq:11} describes various galaxies, it follows that any halo will have a maximum acceleration. A straightforward calculation shows that
the maximal acceleration provided by any DM halo must be $g_{h,{\rm
    max}} = 0.65 g^\dagger = 7.8 \times 10^{-11} {\rm m} \, {\rm
  s}^{-2}$. The existence of this universal maximal
acceleration (UMA) can provide constraints on the surface density of some
of the most common DM profiles in the literature, and in particular on
the properties of scalar field (wave) dark matter (SFDM)
model\cite{Matos:2000ng,Matos:2000ss,Matos:2004rs,Hu:2000ke,Schive:2014dra,Magana:2012xe,Suarez:2013iw,Marsh:2015xka,Hui:2016ltb}.\footnote{A similar approach
  has been pursued in
  Refs.\cite{Karukes:2016eiz,Hayashi:2015maa,Burkert:2015vla,Ogiya:2013qnf,DelPopolo:2012eb,Milgrom:2009bi,Donato:2009ab},
  and their results are in agreement with ours once the appropriate
  conversions between physical quantities are taken into account.}

For purposes of generality, let us assume that the DM density profile
is spherically symmetric and given in the form $\rho(r) = \rho_s
f(r/r_s)$, where $\rho_s$ and $r_s$ are the characteristic density and
scale radius of the profile, respectively, and $f(r/r_s)$ is any given
function of its argument. Notice that in the case of profiles with a
core we expect that $f(0)=1$ and therefore $\rho_s$ is the central
density. The magnitude of the (radial) gravitational acceleration
produced by the DM halo at a radius $r$ can be calculated from $g_h(r)
= G M(r)/r^2$, where $G$ is Newton's constant and $M(r)$ is the
enclosed mass inside a sphere of radius $r$. Then, given the general
form of the density profile $\rho(r)$, the gravitational acceleration
can be written as $g_h(r) = G \mu_{DM} \, \hat{g}_h(\hat{r})$, where
$\hat{g}_h(\hat{r}) = (4\pi/ \hat{r}^{2}) \int \limits^{\hat{r}}_0
f(x) x^2 \, dx$ is a dimensionless quantity, and the radial coordinate
has been normalized to the characteristic radius as $\hat{r} = r/r_s$.

We can see then that the gravitational acceleration at any given
radius is proportional to the DM central surface density, which we define
simply as $\mu_{DM} = \rho_s r_s$. Furthermore, for any density
profile the derived maximal acceleration is given by
\begin{equation}
  \label{eq:10}
  \frac{g_{h,\rm max}}{10^{-11} \, {\rm m} \, {\rm s}^{-2}}  = 0.014
  \left( \frac{\mu_{DM}}{M_\odot {\rm pc}^{-2}} \right) \hat{g}_{h, \rm
    max} \, .
\end{equation}
The value of the dimensionless maximal acceleration $\hat{g}_{h, \rm
  max}$ can be readily calculated for any density profile
$f(\hat{r})$, and then Eq.~\eqref{eq:10} directly becomes a constraint
equation for the DM surface density $\mu_{DM}$. 

We have selected DM density profiles of common use in the literature,
and derived the expected value of their central DM surface density
$\mu_{DM}$ by imposing that each profile satisfies the
UMA Eq.~\eqref{eq:10} at their corresponding point of maximal acceleration
$\hat{g}_{h,\rm max}$. The left hand side of Eq.~\eqref{eq:10} is the
result of a mean behavior of various galaxies, then the derived values of
$\mu_{DM}$ in Table~\ref{tab:table1} will represent the expected
overall behavior that the best-fit parameters of individual galaxies
should follow. Our predicted values are in agreement with
those reported in previous works for the
Burkert\cite{Donato:2009ab,Robles:2012uy,Rodrigues:2017vto},
MultiState SFDM\cite{Martinez-Medina:2014hka}, pseudo-isothermal (PI)\cite{Robles:2012uy} and that of Spano et al\cite{Spano:2007nt} profiles. It can be seen that the standard CDM profile,
also known as the Navarro-Frenk-White (NFW)
profile\cite{Navarro:1996gj}, shows an acceleration that converges to
its maximum at the center. That is, the maximum DM acceleration of the
NFW profile is predicted to happen at the center of each galaxy, which
is also noticed in CDM simulations\cite{Navarro:2016bfs}. In contrast,
we find that for all core profiles the accelerations will reach their
maximal value near their scale radius $r_s$ and then drop to zero for
smaller radii.

\begin{table*}
  \centering
  \caption{Row(1): DM models and their characteristic
    quantities. Row(2) Dimensionless density profiles.  Row(3): Value
    of the dimensionless radius $\hat{r}_{\rm max} = r_{\rm max}/r_s$
    at which the DM acceleration is maximal. Row(4): Maximal
    value of the (dimensionless) radial acceleration  $\hat{g}_{\rm
      max}$. Row(5): The surface density $\mu_{DM} = \rho_s r_s$ (in
    units of $M_{\odot} {\rm pc}^{-2}$) obtained from the
    constraint~\eqref{eq:10}.\label{tab:table1}}
  \begin{tabular}{lcccccc} 
    \hline
    & Burkert & MultiState-SFDM & PI & Spano & WaveDM & NFW \\
    \hline
    $f(r/r_s)$ & $\left[ (1+r/r_s)(1+r^2/r^2_s) \right]^{-1}$ & 
    $\sin^2(r/r_s)/(r/r_s)^2$ &
    $\left(1+r^2/r^2_s \right)^{-1}$ & $\left(1+r^2/r^2_s
                                       \right)^{-3/2}$ &
    $\left(1+r^2/r^2_s \right)^{-8}$ & $\left[ (r/r_s)(1+r/r_s)^2
                                       \right]^{-1}$ \\
    $\hat{r}_{\rm max}$ & $0.96$ & $1.57$ & $1.51$ & $1.03$ & $0.36$ & $0$ \\
    $\hat{g}_{\rm max}$ & $1.59$ & $4.02$ & $2.89$ & $2.19$ & $0.86$ & $2\pi$ \\
    $\mu_{DM}$ & $580$ & $355$ & $369$ & $541$ & $648$ & $89$ \\
    \hline
  \end{tabular}
\end{table*}

Although the UMA obtained from the MDAR is a universal quantity
independent of the DM profile, this is not the case for the surface
density $\mu_{DM}$ or the dimensionless maximal value $\hat{g}_{h, \rm
  max}$, both dependent on the chosen DM profile. Nonetheless, once
$\hat{g}_{h, \rm max}$ is calculated for a given density profile, its
associated value of $\mu_{DM}$ will be fixed for all halos modeled using the same profile, which then implies the correlation of the two free
parameters in the density profile $\rho_s$ and $r_s$; the latter are
allowed to vary from galaxy to galaxy as long as their product $\rho_s
r_s$ remains a constant. If the UMA in the DM is valid independently of the 
baryonic matter content in a galaxy, it implies that all DM profiles in
Table~\ref{tab:table1} have only one free parameter to fit the 
rotation curve of any individual galaxy.

Empirical core profiles have parameters that are not necessarily tied
to fundamental properties of the DM; however, the profile parameters
in models that are theoretically motivated can be related to intrinsic
quantities of the model under study.  One particularly interesting
case that falls in the latter category is that of ultra-light SFDM,
which assumes that the DM particle is a scalar field of very small mass whose quantum properties appear at galactic
scales\cite{Matos:2000ng,Matos:2000ss,Hu:2000ke,Suarez:2013iw,Marsh:2015xka,Hui:2016ltb}. Although
the relativistic theory may be complicated\cite{Hlozek:2014lca,Urena-Lopez:2015gur}, the properties of the halo
density profile are dictated by those of the so-called
Schrodinger-Poisson (SP) system of equations, see\cite{Guzman:2004wj,Guzman:2006yc} and references therein. The
soliton profile actually corresponds to the ground state solution of
the SP system, and we refer to it as the WaveDM profile to distinguish
it from other more general solutions of the SFDM model,e.g., considering multiple states of the field\cite{Robles:2013ftm} whose analytical profile is also included in Table~\ref{tab:table1} (MultiState-SFDM).

\begin{figure*}[htp!]
\includegraphics[width=0.49\textwidth]{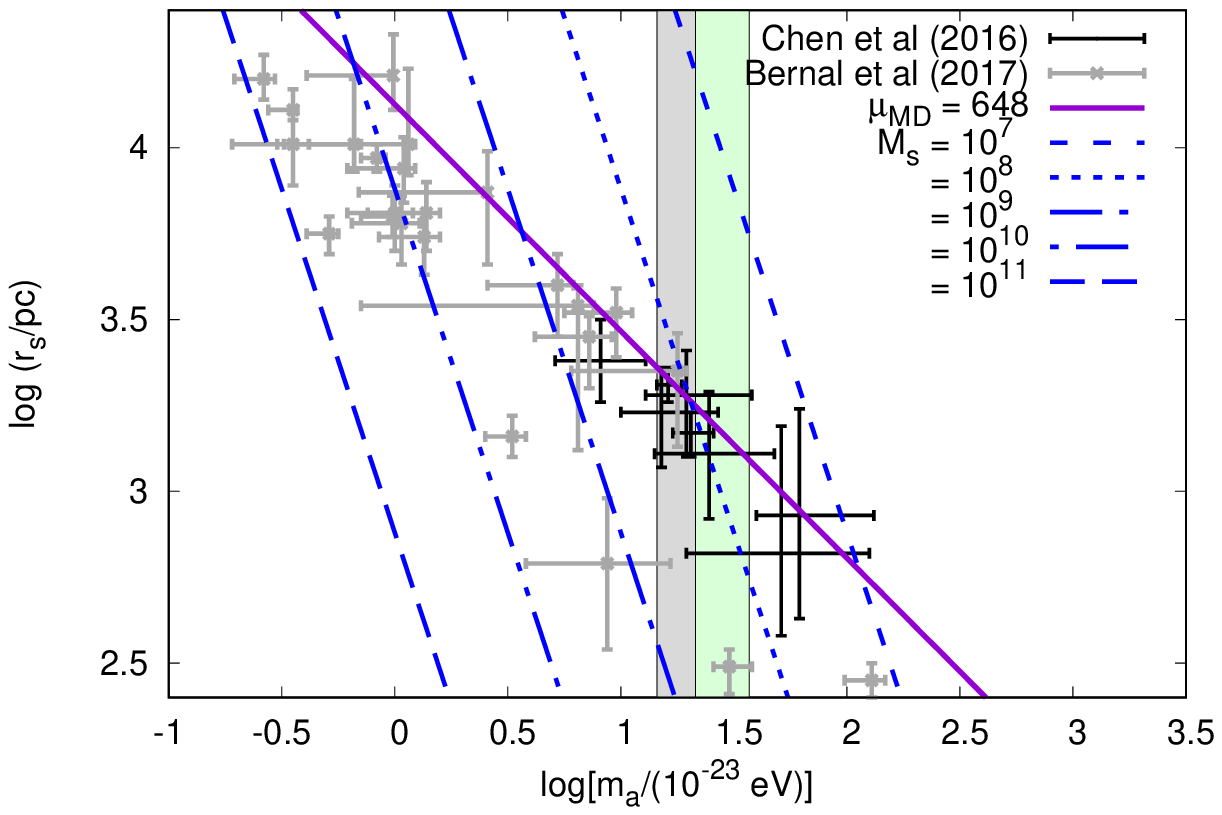}
\includegraphics[width=0.49\textwidth]{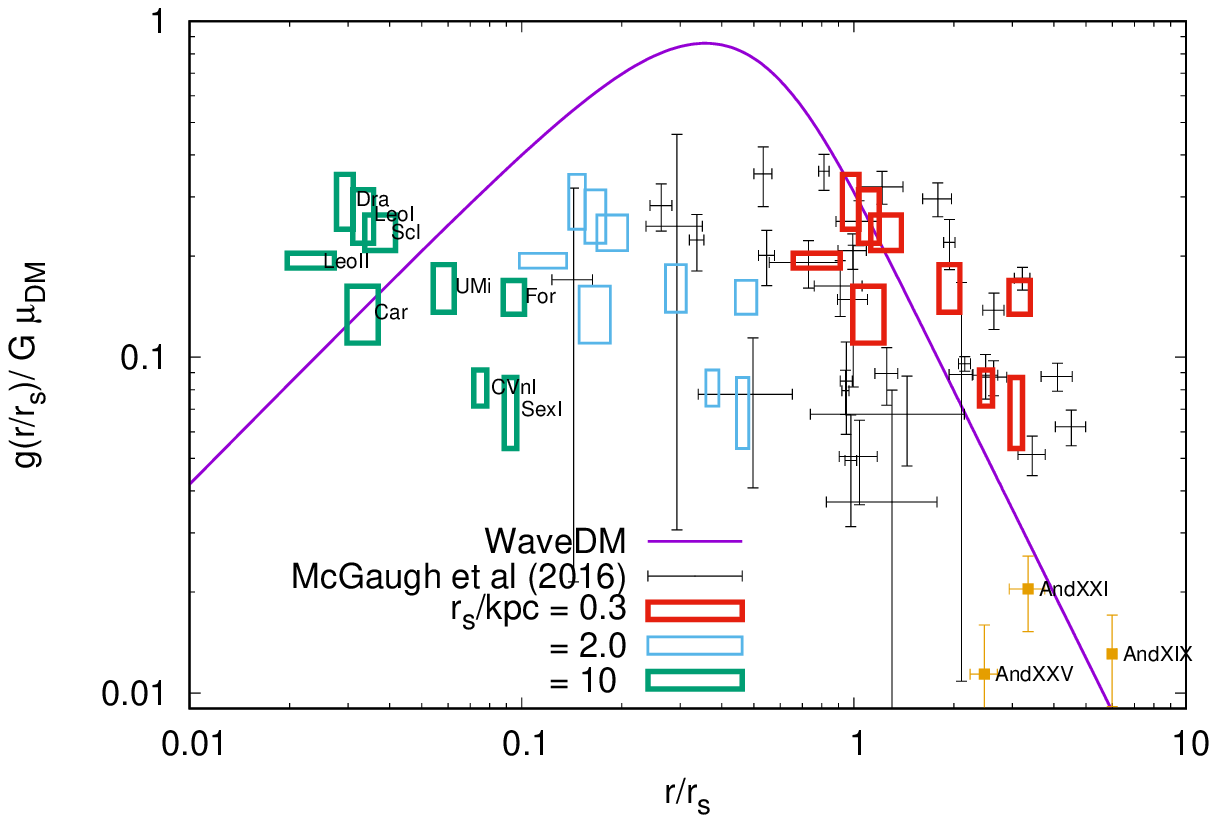}
 \caption{(Left) The correlation of the parameters $m_a$ and $r_s$ of
   the WaveDM profile predicted from the MDAR via Eq.~\eqref{eq:6a}
   (magenta line). The data points correspond to the \emph{individual} fits
   of the WaveDM profile to dwarf galaxies as reported in Chen et
   al\cite{Chen:2016unw} and in Bernal et al\cite{Bernal:2017oih} in
   which both $m_a$ and $r_s$ were treated as independent. Notice that
   the resultant points follow the trend of the correlation dictated by
   the constraint~\eqref{eq:6a} with $\mu_{DM} = 648 \, M_{\odot} {\rm
     pc}^{-2}$. Performing the fits only to the classical MW dSphs, as
   done in\cite{Chen:2016unw,Gonzales-Morales:2016mkl}, results in an
   apparent narrow range of the boson mass due to the selection of the
   sample. In Ref.\cite{Chen:2016unw} the estimated boson mass was
   $m_a = 1.79 \times 10^{-22} {\rm eV/c^2}$ (grey shaded region),
   whereas in Ref.\cite{Gonzales-Morales:2016mkl} the obtained value
   was $m_a = 2.4 \times 10^{-22} {\rm eV/c^2}$ (green shaded
   region). From Eq.~\eqref{eq:6a} the scale radius for the
   aforementioned two values of the boson mass are $r_s = 2.1 \, {\rm
     kpc}$ and $r_s = 1.6 \, {\rm kpc}$, respectively. The blue lines
   represent the values of constant soliton mass $M_s$ enclosed within
   the radius $r_s$.\footnote{The soliton mass is calculated from the
     formula $M_s = 7.7 \times 10^{13} M_\odot (m_a/10^{-23} {\rm
       eV})^{-2} (r_s/{\rm pc})^{-1}$, see for
     instance\cite{Guzman:2004wj,Marsh:2015wka,Chen:2016unw}.} If the boson mass
   $m_a$ is allowed to freely vary, the sample of galaxies suggests a
   variation in $m_a$ ($M_s$) by two (four) orders of
   magnitude. (Right) Normalized gravitational acceleration $\hat{g} =
   g/(G \, \mu_{DM})$ inferred from the classical MW dSphs according
   to the determination of the mass inside the half-light radius
   $r_{1/2}$ reported in\cite{Fattahi:2016nld} (represented by the
   errorboxes), compared to the
   expected theoretical curve of WaveDM. Three values of $r_s$ were
   chosen for the normalization of the half-light radii of each
   galaxy. Notice that the MW data seem to prefer small values of
   $r_s$ (red errorboxes), for which the data seem to follow the
   downward trend of the curve, whereas values of $r_s \geq 2 \, {\rm
     kpc}$ (blue and green errorboxes) look inappropriate. Also
   plotted are all the MW and Andromeda satellites with high-quality
   data from\cite{McGaugh:2016oct} using the same scale radius
   $r_s=0.3 \, {\rm kpc}$. In particular, AndXIX, AndXXI and AndXXV,
   previously considered outliers and regarded as low mass for their
   size in Ref.\cite{Collins:2013eek}, seem to agree well with the
   downward trend of the theoretical curve.}
\label{fig:1}
\end{figure*}

In empirical profiles the two parameters $\rho_s$ and $r_s$ are
treated as independent, and they are not linked to any particular DM
property. For the WaveDM profile, however, the parameters $\rho_s$ and
$r_s$ are predicted to have the following scaling property: $\rho_s =
\lambda^4 m^2_a \, m^2_{\rm Pl}/4\pi$ and $r_s = (0.23 \, \lambda \,
m_a)^{-1}$, where $m_{\rm Pl}$ is the Planck mass, $m_a$ is the mass
of the boson particle and $\lambda$ is a scaling
parameter\cite{Guzman:2004wj,Marsh:2015wka}. By the elimination of the scaling
parameter $\lambda$, we then find the following expression for the
surface density $\mu_{DM}$ in terms of the mass $m_a$ and the soliton
radius $r_s$,
\begin{equation}
  \label{eq:6a}
  \left( \frac{r_s}{\rm pc} \right)^{-3} \left( \frac{m_a}{10^{-23}
      {\rm eV}} \right)^{-2} = 4.1 \times 10^{-15} \left(
    \frac{\mu_{DM}}{M_\odot {\rm pc}^{-2}} \right) \, .
\end{equation}
The existence of a universal value of the surface density, namely
$\mu_{DM} = 648 \, M_{\odot} \, {\rm pc}^{-2}$ (see
Table~\ref{tab:table1}), implies, similarly to other core
profiles, a close correlation between $m_a$ and $r_s$.\footnote{Notice
  that the universality of $\mu_{DM}$ in Eq.~\eqref{eq:6a} implies for
  the WaveDM profile the correlation $m_a \propto r^{-3/2}_s$, which is
  also reported in Fig. 2 of Ref.\cite{Chen:2016unw}, although it is
  erroneously attributed there to a constant core density
  $\rho_s$. See also Fig. 7 in Ref.\cite{Bernal:2017oih}.} Moreover,
within the WaveDM paradigm the boson mass $m_a$ is a fundamental
physical parameter and as such it cannot vary from galaxy to
galaxy. If we now consider the universality of the surface density
implied by the MDAR, Eq.~\eqref{eq:6a} must also fix the value of the
scale radius $r_s$. In contrast to other empirical profiles, we then
find that the MDAR in Eq.~\eqref{eq:11} ultimately implies the
existence of an universal soliton (core) profile within the WaveDM
paradigm.

Notably, if we neglect the assumption that the boson mass $m_a$ is
fundamental and treat it as a free parameter, then the best-fit
parameters modeling the rotation curves of individual galaxies are
expected to satisfy Eq.~\eqref{eq:6a}. However, this fitting procedure
will inevitably lead to a large dispersion in the mass which is simply
a consequence of the diversity in galaxy sizes. Thus, as long as $m_a$
is treated as a free fitting parameter to describe a diverse sample of
galaxies, we cannot derive a meaningful constraint of its value. The
left panel in Fig.~\ref{fig:1} illustrates this point, where we show
the values of individual galaxy fits obtained in previous
works\cite{Chen:2016unw,Gonzales-Morales:2016mkl,Bernal:2017oih} in
which $m_a$ and $r_s$ were treated as independent fitting
parameters. In general, the fits lie closely along the line of
constant surface density indicated by Eq.~\eqref{eq:6a}, which is the
expected behavior from the universality of the MDAR. The large scatter
in $m_a$ is the reflection that the fitting method in
\cite{Chen:2016unw,Gonzales-Morales:2016mkl,Bernal:2017oih} cannot
provide a reliable determination of the boson mass $m_a$, and that the
most they can do is to test the reliability of the
constraint~\eqref{eq:6a}.

Nonetheless, as we shall show, Eq.~\eqref{eq:6a}, along with the
assumptions that $m_a$ is constant and that all halos are described by
the WaveDM profile, are enough to derive a simple estimate of the
boson mass, we only require an independent estimate of $r_s$. 
We will use data from dwarf spheroidal (dSph) galaxies for our estimate. Being
DM dominated systems, we expect the properties of dSphs to be
similar, albeit with some possible scatter
associated to their formation histories\cite{Fitts:2016nlc}.

In the right panel of Fig.~\ref{fig:1} we show the (theoretical) gravity profile
$\hat{g}(r/r_s)$ of the WaveDM model, together with the acceleration
values reported in Ref.\cite{Fattahi:2016nld} for the Milky Way (MW)
dSphs. The data correspond to the enclosed mass $M_{1/2}$ within the
half-light radius $r_{1/2}$, which were then converted to a
gravitational acceleration at the same radius through the equation $g
(r_{1/2}) = GM_{1/2}/r^2_{1/2}$. The latter was then normalized as $g/G
\mu_{DM}$ for a proper comparison with $\hat{g}$ using the value of
$\mu_{DM} = 648 \, M_\odot \, {\rm pc}^{-2}$ found from the MDAR for the
WaveDM profile. Hence, the only free parameter to adjust is $r_s$,
which we use to normalize the half-light radius $r_{1/2}$ of each of
the classical MW dSphs. We chose three different values, namely
$r_s/{\rm kpc = 0.3, 2, 10}$, which then correspond to three relative
positions of the data points with respect to the theoretical curve. It
can be seen that the best option is $r_s = 0.3 \, {\rm kpc}$ (red
errorboxes in Fig.~\ref{fig:1}), which puts the data points on the
right hand side of the point of maximal acceleration, where they even
seem to follow the downward trend of the theoretical curve at
large radii. 

Using the same value of $r_s = 0.3 \, {\rm kpc}$ we scale a bigger
sample of satellite galaxies\cite{Collins:2013eek,McGaugh:2016oct}, 
in Fig.~\ref{fig:1} we include only the dwarf galaxies with high-quality
observations selected in\cite{McGaugh:2016oct}. Surprisingly, we find
that the data  follows the theoretical curve reasonably well after the
point of maximal acceleration, the largest scatter coming from the
galaxies with the large observational uncertainties.  Even more, we see
that satellites AndXIX, and AndXXI and AndXXV, which were labeled as
outliers in the study of Ref.\cite{Collins:2013eek}, mostly because of
their low mass as compared to their size, seem to vindicate the trend
of the theoretical curve of the WaveDM profile at the lowest values of
the gravitational acceleration.

For our estimated $r_s = 0.3 \, {\rm kpc}$, the constraint
equation~\eqref{eq:6a} implies the boson mass $m_a = 1.2 \times
10^{-21} \, {\rm eV}$, and the soliton mass $M_s = 1.8 \times 10^7 \,
M_\odot$, which is consistent with the uniformity of the mass
estimates within $300 \, {\rm pc}$ made in
Ref.\cite{Strigari:2008ib}. Notice that the boson mass is somehow
unexpected, as it is much larger than the values commonly reported in
the literature for dwarf
galaxies\cite{Schive:2014dra,Chen:2016unw,Marsh:2015wka,Gonzales-Morales:2016mkl}. However, this new and larger value of
the boson mass is constant for  all halos, as demanded by the
hypothesis of the SFDM model, and avoids the stringent constraints
coming from cosmological observations\cite{Hui:2016ltb,Corasaniti:2016epp,Schive:2015kza,Menci:2017bfs,Banik:2017ygz,Gonzales-Morales:2016mkl,Sarkar:2017vls}.

Because the MDAR implies a single value of the scale radius $r_s$ in
the WaveDM profile, there should be an universal soliton profile
present in all galaxy halos, but due to the diversity of galaxy sizes
this implies the WaveDM profile alone will prove unable to describe
all DM halos, in particular large
ones\cite{Martinez-Medina:2014hka}. This does not rule out the SFDM
model, but it does rule out the possibility that the soliton profile
represents all the DM halo; hence, a more general profile than the
WaveDM profile is required that extends to larger radii.

One natural approach that has been proposed is to account for the
superposition of excited states in the SFDM halo, which mostly affect
the outer parts of the halo leaving the characteristic imprint of
\emph{wiggles or oscillations} in the density and rotation curve
profiles (see the MultiState profile in
Table~\ref{tab:table1})\cite{Robles:2013ftm,Martinez-Medina:2014hka,Matos:2007zza}. A
more ad hoc approach deals with smoothly matching the soliton to a NFW
profile describing the outer part of the halo which adds a second
parameter to the fitting density
profile\cite{Marsh:2015wka,Gonzales-Morales:2016mkl,Bernal:2017oih},
this is motivated by the results of numerical simulations of the SP
system\cite{Schive:2014dra,Schive:2014hza,Schwabe:2016rze,Du:2016zcv,Du:2016aik}.

We have performed numerical calculations on this more general
profile (following the prescription in Ref.\cite{Bernal:2017oih}) that
models the full SFDM halo and applied to it the MDAR
constraint Eq.~\eqref{eq:10}. The resultant SFDM surface density is found
to lie in the range $\mu_{DM} = (575-648) \, M_\odot {\rm pc}^{-2}$; 
these values are of the same order of magnitude as in the case of the
soliton profile alone, and they would similarly imply that $r_s\sim
0.3 \, {\rm kpc}$. We find that considering this general profile leaves the main results
about the unique soliton and the estimated boson mass roughly
unchanged and opens the possibility of fitting a diverse sample of
galaxies to their outermost radii with a single boson mass. At the
same time it also strengthens the possibility of an universal soliton
profile with a total mass of $10^7 \, M_\odot$ in the center of all
galaxy halos.\cite{Strigari:2008ib,Okayasu:2016oyy}.

Summarizing, we have shown that the MDAR implies in general a
universal value of the surface density for any given DM density
profile, which translates into a strong correlation between their
central density $\rho_s$ and scale radius $r_s$. We
have explored the consequences of such correlation in the case of the
SFDM model, which leads to the conclusion that all galaxy halos should
have a similar central core structure, formed due to the Heisenberg
uncertainty principle. Comparison with data of satellite galaxies of
the MW and Andromeda suggests a boson mass in the SFDM model of
$10^{-21} \, {\rm eV}/c^2$, which is in agreement with current
cosmological and astrophysical constraints, while still having a distinguishable history of structure formation and halo density
distribution from the standard CDM model\cite{Schive:2015kza,Du:2016zcv,Corasaniti:2016epp}.

\begin{acknowledgments}
V.H.R. acknowledges CONACyT M\'exico for financial support. This work
was partially supported by CONACyT M\'exico under grants
CB-2014-01, no. 240512, CB-2011, no. 166212, Xiuhcoatl and Abacus
clusters at Cinvestav, 49865-F and I0101/131/07 C-234/07 of the
Instituto Avanzado de Cosmologia collaboration. LAU-L acknowledges
support by the Programa para el Desarrollo Profesional
Docente; Direcci\'on de Apoyo a la Investigaci\'on y al Posgrado,
Universidad de Guanajuato, research Grant No. 732/2016; CONACyT
M\'exico under Grants No. 167335 and No. 179881; and the Fundaci\'on
Marcos Moshinsky.
\end{acknowledgments}

\bibliography{centro}

\end{document}